# Pairing symmetry in infinite-layer nickelate superconductors

L. E. Chow,[1,#] S. Kunniniyil Sudheesh,[1,2,#] Z. Y. Luo,[1] P. Nandi,[1] T. Heil,[3] J. Deuschle,[3]

S. W. Zeng,[1] Z. T. Zhang,[1] S. Prakash,[1] X. M. Du,[1] Z. S. Lim,[1] Peter A. van Aken,[3]

Elbert E. M. Chia,[2,*] A. Ariando,[1,*]

[1]Department of Physics, Faculty of Science, National University of Singapore, Singapore 117551, Singapore

[2]Division of Physics and Applied Physics, School of Physical and Mathematical Sciences, Nanyang Technological University, 21 Nanyang Link, Singapore 637371, Singapore

[3]Max Planck Institute for Solid State Research, 70569 Stuttgart, Germany

[#]The authors contributed equally to this work

*To whom correspondence should be addressed: ariando@nus.edu.sg; elbertchia@ntu.edu.sg

## Abstract

The superconducting infinite-layer nickelate family has risen as a promising platform for revealing the mechanism of high-temperature superconductivity. However, its challenging material synthesis has obscured effort in understanding the nature of its ground state and low-lying excitations, which is a prerequisite for identifying the origin of the Cooper pairing in high-temperature superconductors. In particular, the superconducting gap symmetry of nickelates has hardly been investigated and remains controversial. Here, we report the pairing symmetry of the infinite-layer nickelates determined by London penetration depth measurements in neodymium-based $(Nd,Sr)NiO_2$ and lanthanide-based $(La,Ca)NiO_2$ thin films of high crystallinity. A rare-earth-specific order parameter is observed. While the lanthanide nickelates follow dirty line-node behaviour, the neodymium-counterpart exhibits nodeless order parameters such as the $(d + is)$-wave. In contrast to the cuprates, our results suggest that the superconducting order parameter in nickelates is beyond a single $d_{x^2-y^2}$-wave gap. Furthermore, the superfluid density shows a long tail near the superconducting transition temperature which is consistent with the emergence of a two-dimensional to three-dimensional crossover in the superconducting state. These observations challenge the early theoretical framework and propel further experimental and theoretical interests in the pairing nature of the infinite-layer nickelate family.

## Main Text

Since the discovery of zero electrical resistance and perfect diamagnetism in superconductors, understanding the mechanism of superconductivity and manipulating these phenomena for room-temperature application has been one of the longstanding challenges in physics. The success of Bardeen-Cooper-Schriefer (BCS) theory, where the electron-phonon interaction explains the origin of conventional superconductivity,[1] also leads to a calculation result which bottlenecks the maximum possible superconducting transition temperature to a few tens of kelvin,[2] puts an end to the hope of achieving high-temperature superconductivity. However, the later and unexpected discovery of high superconducting transition temperature (high-$T_c$) in the cuprates at ambient pressure that falls outside the BCS paradigm brought about the dawn of high-$T_c$ superconductivity and once again revived the hope of achieving room-temperature superconductivity[3]. In contrast to BCS superconductors, which exhibit isotropic *s*-wave pairing symmetry, the high-$T_c$ superconductors are typically *d*-wave in nature with nodes in the superconducting order parameter[4]. Since then, physicists have been aspiring to understand the mechanism of high-$T_c$ superconductivity by seeking and studying compounds that are isostructural to the cuprates[5]. The first experimental possibility was demonstrated on $Sr_2RuO_4$, which replaces the Cu atom in the Cu-$O_2$ plane with another transition element Ru[6], which later was found to have a superconducting pairing symmetry that is different from the *d*-wave symmetry of the cuprates[7]. Another option is the substitution of $Cu^{2+}$ with $Ni^{1+}$, which results in nickelates that retain a similar $3d^9$ electronic structure that is envisioned to play a crucial role in unconventional pairing in cuprates[5,8–10]. This decades-long effort[5,11–13] has been rewarded recently with the experimental confirmation of superconductivity in the infinite-layer nickelate thin films[8,14–19].

Since then, attention has been drawn to nickelates to understand the origin of unconventional superconductivity by comparing the similarities and differences between nickelates and cuprates[10,11,20,21]. For example, recent resonant inelastic X-ray measurements have confirmed the presence of charge order and antiferromagnetic order in infinite-layer nickelate with a lower magnon energy than in the cuprates[22–24]. However, the fabrication of the nickelate superconductor has been shown to be extremely challenging, while the thin film form of nickelates that is subjected to strain and film-substrate interface effect has added complexity in understanding the system in contrast to typical bulk-form superconductors[15,25–28]. Important questions, such as the pairing symmetry of nickelates, remain controversial and unanswered. Considering the isostructural and isoelectronic nature of nickelates to cuprates, theoretical calculations have supported the notion that the nickelates have similar gap symmetry as of cuprates[29,30], a dominant $d_{x^2-y^2}$-wave pairing despite its multiorbital nature[20,30–32]. Other propositions include (1) a dominant $d_{xy}$-wave gap[33]; (2) a two-gap model $d_{z^2}$-wave + $d_{x^2-y^2}$-wave for neodymium (Nd) and praseodymium (Pr) based nickelates, and single $d_{x^2-y^2}$-wave in the lanthanum (La) based counterpart[34]; (3) $s$-wave and $(d + is)$-wave gap in the nickelates if the hopping $t_c/K$ is sufficiently small as compared to the Kondo coupling, in contrast to a dominant $d$-wave gap for a large $t_c/K$, that is described in the $t - J - K$ model which accounted the antiferromagnetic exchange and Kondo coupling[35]. A previous experimental study on a $Nd_{0.8}Sr_{0.2}NiO_2$ thin film with single-particle tunnelling spectroscopy shows a mixture of $d$-wave and $s$-wave signals at different locations of the film surface, which is prone to the formation of nonstoichiometric or secondary phases, thereby masking the true pairing symmetry of the superconducting gap[18,36]. In addition, theoretical studies have debated on the role of the $4f$ magnetism in superconductivity found in the Nd-based and Pr-based infinite-layer nickelates, while experimentally, superconductivity in La-based nickelate with an empty $4f$ orbital could not be achieved for two decades, until lately[37,38], despite it being the earliest studied rare-earth nickelates[8,39–41]. An open question is whether the La- nickelates and Nd-nickelates have different symmetries in the superconducting order parameter,[42] and consequently, what the pairing mechanism of the infinite-layer nickelate superconductor family is.

Here we report a successful growth of superconducting Nd-based and La-based infinite-layer nickelate thin films of high crystallinity and study their pairing symmetry through the London penetration depth measurement using a tunnel diode oscillator technique[43,44]. Our results (1) suggest that a single $d$-wave gap is unlikely to be the complete picture of the pairing symmetry in the nickelate family; (2) demonstrate distinct pairing symmetries between Nd- and La-nickelates, where Nd-nickelate likely hosts nodeless multigap pairing such as $(d + is)$-wave while La-nickelate hosts anisotropic nodal order parameters such as $(d + s)$-wave or $(d + p)$-wave; (3) propose a crossover from two-dimensional to three-dimensional superconducting states in the infinite-layer nickelate thin-film which is consistent with the recent angular dependent upper critical field study.[45]

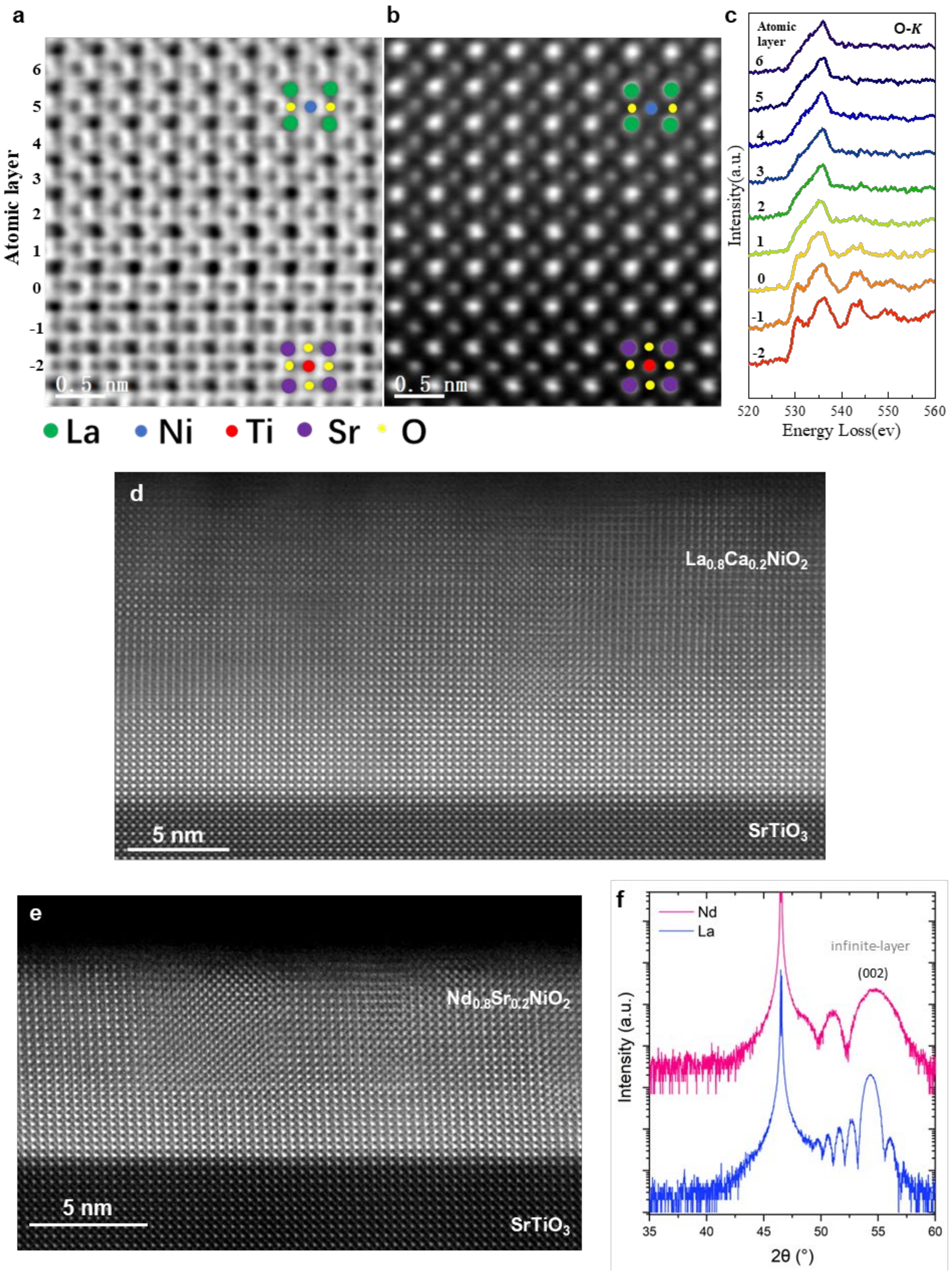

**Figure 1: Structural characterization of the infinite-layer nickelate thin films.** (**a-c**) Layer-resolved annular bright-field (**a**), high-angle annular dark-field (**b**), and electron energy loss spectroscopy at O $K$ edge (**c**) near the film-substrate interface of #L1. (**d-e**) Cross-sectional high-angle annular dark-field STEM images of the superconducting lanthanide- (**d**) and neodymium- (**e**)

infinite-layer nickelate thin film samples #L1 and #N1. (f) X-ray diffraction $\theta - 2\theta$ symmetric scan patterns at the (002) peak of the infinite-layer neodymium (Nd) and lanthanide (La) nickelate thin films. Laue fringes are visible.

## Structural information

Figures 1(a-e) show the cross-sectional scanning transmission electron microscopy and spectroscopy characterization of the infinite-layer nickelate thin films. Figure 1f shows the X-ray diffraction (XRD) $\theta - 2\theta$ symmetric scan of the infinite-layer nickelate thin films. High crystallinity is verified by the observation of a clear Laue fringes pattern in the vicinity of the (002) peak for the reduced infinite-layer phase of both Nd- and La-nickelate thin films. No defect phase was observed. The c-axis lattice constants are calculated to be 3.38 Å and 3.35 Å for the La and Nd-based samples, respectively. Large monocrystalline areas can be observed in the cross-sectional STEM high-angle annular dark-field (HAADF) images of La- and Nd-based infinite-layer nickelate thin film samples # L1 and #N1, respectively, as shown in Figure 1(d-e). For the Nd-based nickelate thin film sample #N1 (Figure 1e), small areas of Ruddlesden-Popper (RP) stacking faults can be observed, as reported in the previous studies[46]. The crystal and electronic structure of the infinite-layer phase is further confirmed by the layer-resolved electron energy loss spectroscopy (EELS) and annular bright-field (ABF) images (Figure 1(a-c)) which show the absence of apical oxygen in the reduced nickelate films.

## Transport and magnetic properties

Figure 2a shows the temperature-dependent resistivity of the infinite-layer nickelate thin films. On top of high crystallinity, we observed a clear Meissner effect in the infinite-layer thin-films, represented by the sharp onset and large superconducting volume (from $\chi_V$) seen in the $M - T$ curve and a linear negative slope in the $M - H$ curve. Figure 2b shows the temperature-dependent zero-field cooling (ZFC) volume susceptibility $\chi_V$ measured at a magnetic field applied in the out-of-plane direction. A nearly 100% superconducting volume fraction with $\chi_V \rightarrow -1$ can be seen at 2 K. From the negative slope in the $M - H$ curve shown in Figure S3, we extracted the demagnetization factor and the lower critical field after demagnetization factor correction to be $H_{c1}(T = 0\text{K}) \approx 79\ \text{Oe}$. Figure 2c shows the temperature-dependent Hall coefficient $R_H$ of the infinite-layer thin films. In Nd- nickelate samples, a transition in the dominant charge carrier from electron at high temperature to hole carrier at low temperature is observed, suggesting the dominant role of the $d_{x^2-y^2}$ hole band in the superconducting state at optimal doping ($x = 0.2$). On the other hand, La-based samples show only a negative sign Hall coefficient, even at the lowest temperature.

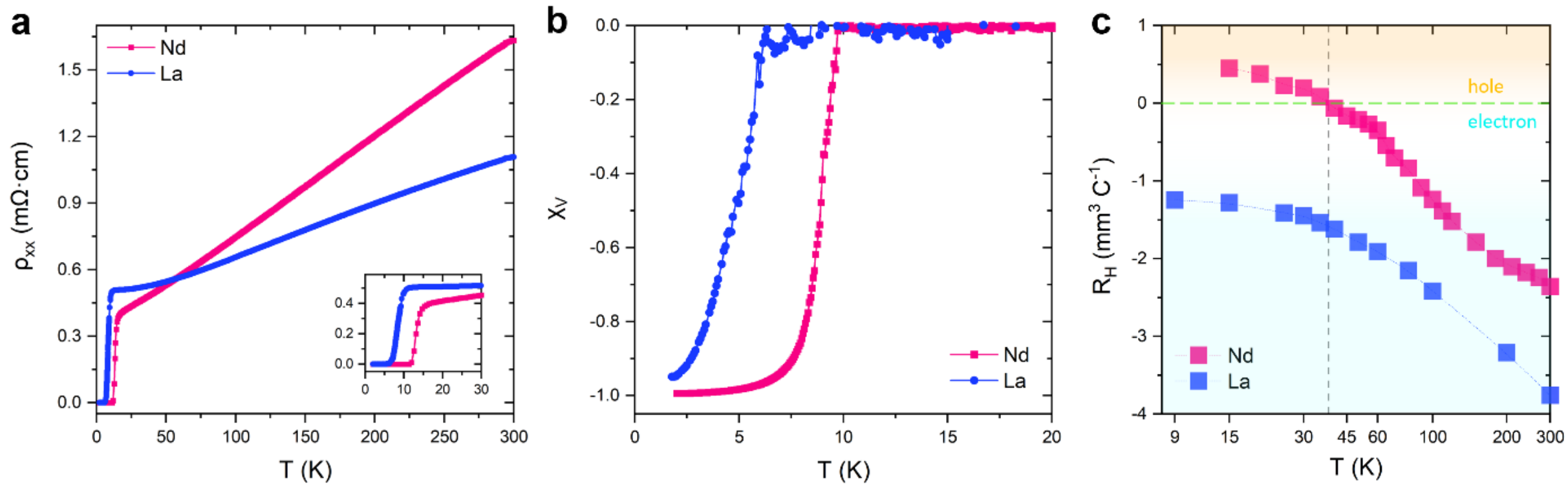

**Figure 2: Transport and magnetic properties of the infinite-layer nickelate thin films.** **(a)** Temperature dependence of linear resistivity $\rho_{xx}$. **(b)** Temperature dependence of zero-field cooling (ZFC) volume susceptibility $\chi_V$ in S.I. unit, measured at $H \parallel c$ =10 Oe, where $\chi_V = -1$ below the onset of the superconducting transition $T_{c,onset}$ indicates a perfect diamagnetic state. **(c)** Temperature dependence of Hall coefficient $R_H$. A dominant hole charge carrier is present in the Nd-nickelate below 38 K (grey line), while electron is the dominant charge carrier down to 9 K in the La-nickelate thin film.

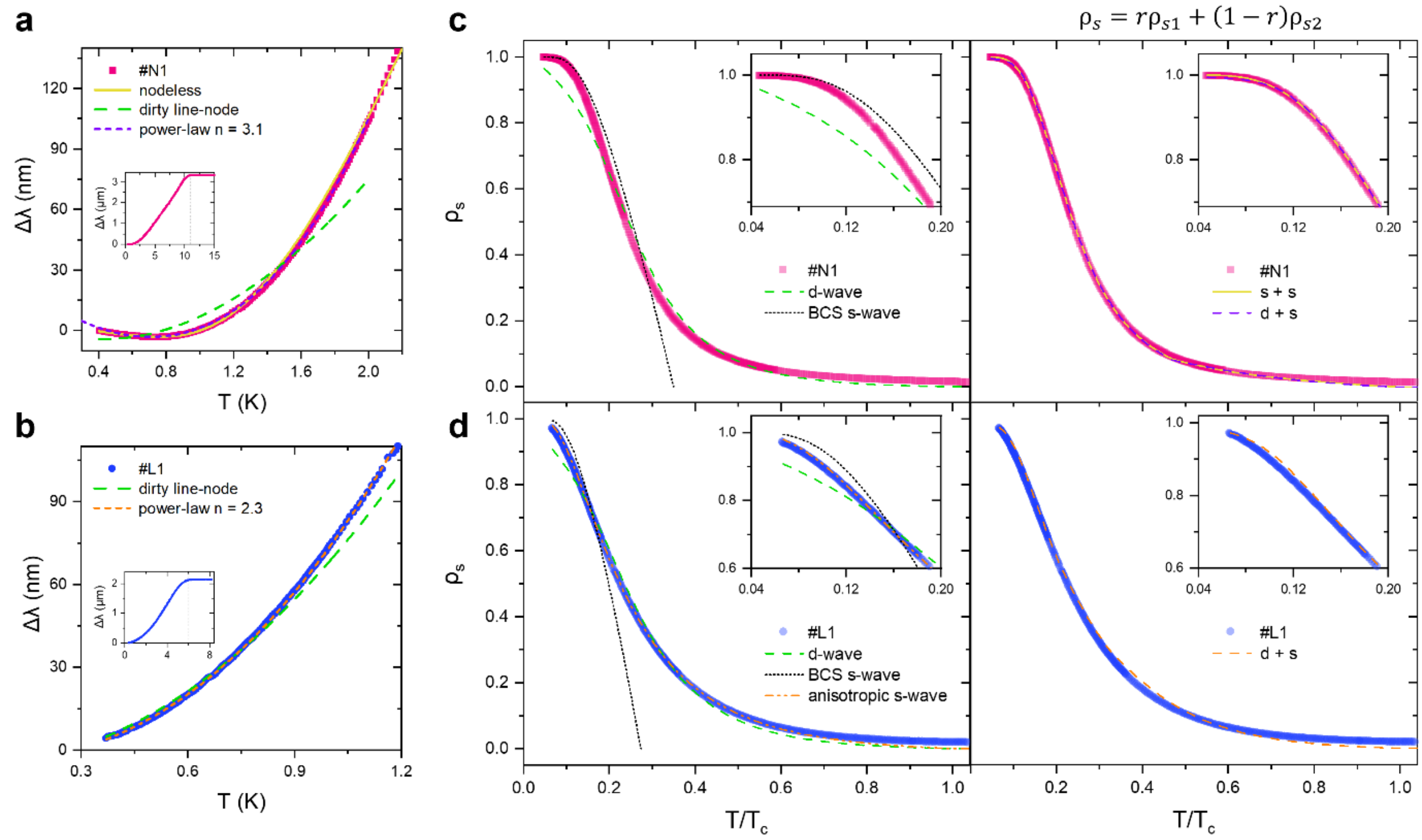

**Figure 3: London penetration depth and superfluid density in the Nd- and La- infinite-layer nickelate thin films. (a-b)** Low-temperature fit of $\Delta\lambda$ for Nd- (#N1) and La- (#L1) infinite-layer nickelate. Insets show the full transition. The onset of transition, $T_c$ is found to be $T_c = 11\text{K}$ for #N1 and $T_c = 6\text{K}$ for #L1, which are close to $T_{c,0}$ in resistivity. For Nd-nickelate (**a**), a low-temperature upturn is observed at < 0.7 K (see Figure S1a). We attribute the upturn to the paramagnetic $Nd^{3+}$ ions, which is supported by the absence of a similar upturn in La-nickelate (**b**). $\Delta\lambda(T \ll T_c)$ is fitted to the nodeless exponential equation and power-law equation ($1 \leq n \leq 2$ for dirty line-node vs free $n$). Exponential equation traces Nd-data better; power-law equation with $n = 2.3$ best traces La-data. Best-fit parameters are listed in Table S1. **(c-d)** Global fit of the normalized superfluid density $\rho_S$ as a function of normalized temperature $T/T_c$ for #N1 (**c**) and #L1 (**d**). Insets show the enlarged view at $T/T_c$ < 0.2. $\rho_S$ data is fitted to theoretical $\rho_S$ of single $d$-wave, BCS $s$-wave with variable $T_c$, anisotropic $s$-wave gap, and two gaps of different $T_c$. All the best-fitted parameters are listed in Table S2.

## London penetration depth analysis

Figures 3(a-b) show the temperature dependence of the change in the in-plane London penetration depth $\Delta\lambda_{ab}$ (noted as $\Delta\lambda$ below) measured on neodymium Nd-based (#N1) and lanthanide La-based (#L1) samples. The in-plane penetration depth is calculated from the frequency shift $\Delta f(T)$ measured in a tunnel-diode-oscillator setup, which we discussed in detail in the **Methods** section. In the low-temperature regime, Figure 3a and Figure S1a show that an upturn in $\Delta\lambda(T)$ is present in the Nd-based nickelate superconductor but absent in the La-based nickelate superconductor (Figure 3b). Hence, such upturn is unlikely to be extrinsic, and we attribute its origin to the paramagnetic $Nd^{3+}$ ions. Similar upturn features in low-temperature $\Delta\lambda(T)$ have been observed in other classes of superconductors that contain the $Nd^{3+}$ ions, including the cuprates $Nd_{2-x}Ce_xCuO_4$ and FeAs-based $RFeAsO_{0.9}F_{0.1}$ (R=Nd)[47,48]. This paramagnetic signal modifies the penetration depth measured using a tunnel diode oscillator as $\lambda_{meas}(T) = \sqrt{\mu(T)}\,\lambda_{ab}(T)$, where $\lambda_{ab}(\mathrm{T})$ is the in-plane penetration depth and $\mu(T)$ is the magnetic permeability whose divergent behavior at low-temperature can be accounted for by the Curie-Weiss law of paramagnetism $\mu(T) = 1 + \chi(T) = 1 + \frac{C}{T+\theta}$ where $C$ is the Curie constant and $\theta$ is the Curie-Weiss temperature[47].

The low-temperature penetration depth $\Delta\lambda(T \ll T_c)$ is very sensitive to the presence of nodes in the superconducting gap, and the temperature dependence at this regime is an indication of the pairing symmetry. An exponential behaviour of the form $\Delta\lambda(T) \propto \left[\sqrt{\frac{\pi\Delta(0)}{2k_B T}}\exp\left(-\frac{\Delta(0)}{k_B T}\right)\right]$ is expected for a fully-gapped isotropic $s$-wave gap, where $\Delta(0)$ is the gap magnitude at $\mathrm{T} = 0$ K, while a linear temperature dependence is expected for line nodes, such as clean $d$-wave in a quasi-2D Fermi surface[4,47,49]. With increased impurities scattering, the penetration depth at low-temperature $\Delta\lambda(\mathrm{T} \ll \mathrm{T}_c)$ for a nodal $d$-wave superconductor is expected to change from linear in the clean limit to quadratic in the dirty limit[50]. We fitted $\Delta\lambda(T \ll T_c)$ to the nodeless BCS $s$-wave model of exponential behavior and a power-law equation $\Delta\lambda(T) \propto T^n$ with $\Delta(0)$ and $n$ as the respective fitting parameters. We accounted for the paramagnetic contribution in the fitting of low-temperature dependence of $\Delta\lambda(T)$ by including the $\sqrt{\mu(T)} = \sqrt{1 + \frac{C}{T+\theta}}$ factor in the fitting equations with $C$ and θ as fitting parameters. The fitting results are shown in Figure 3(a-b) and Table S1.

To ensure the reproducibility and robustness of our data and analysis, we also measured the penetration depth on another set of samples, noted as #N2 (Nd) and #L2 (La); the results are in Figure 4. A dirty line-node gap can fit the data reasonably well at the low-temperature limit for the La-nickelate but fails to describe the behaviour of Nd-nickelate. A similar observation can be made by fitting the normalized superfluid density $\rho_S = [\lambda(0)/\lambda(T)]^2$ (which is proportional to the phase stiffness $[1/\lambda(T)]^2$) at the low-temperature limit, as shown in Figure 5. For the Nd-nickelate, a nodeless exponential behaviour fits both the $\Delta\lambda(T \ll T_c)$ and $\rho_S(T \ll T_c)$ well. However, the gap magnitudes $\Delta(0)$ obtained from the BCS exponential fit is much smaller than the BCS weak-coupling value of $\Delta(0) = 1.76\, k_B T_c$. This implies the presence of multiple gaps, in which case, the data in the low-temperature regime will be dominated by the gap with the smallest magnitude. Power-law fits the $\Delta\lambda(T \ll T_c)$ data well with an exponent $n > 3$ for Nd-based samples and a smaller $n \geq 2$ for La-based samples. For Nd-nickelate samples, fitting with the power-law equation to a lower temperature range will result in obtaining best-fit power $n > 4$, which is considered to be equivalent to an exponential $s$-wave fitting. On the other hand, for La-nickelate, even though the power-law exponent increases if fit to a lower temperature range, the power-law exponent is retained at $2 \leq n < 4$ down to the lowest temperature range. The distinctions between Nd-nickelate and La-nickelate, and the trend in the best-fit power $n$, are consistent over a wide range of $\lambda(0)$ values, as shown in Figure 6, which suggest the reliability of the results.

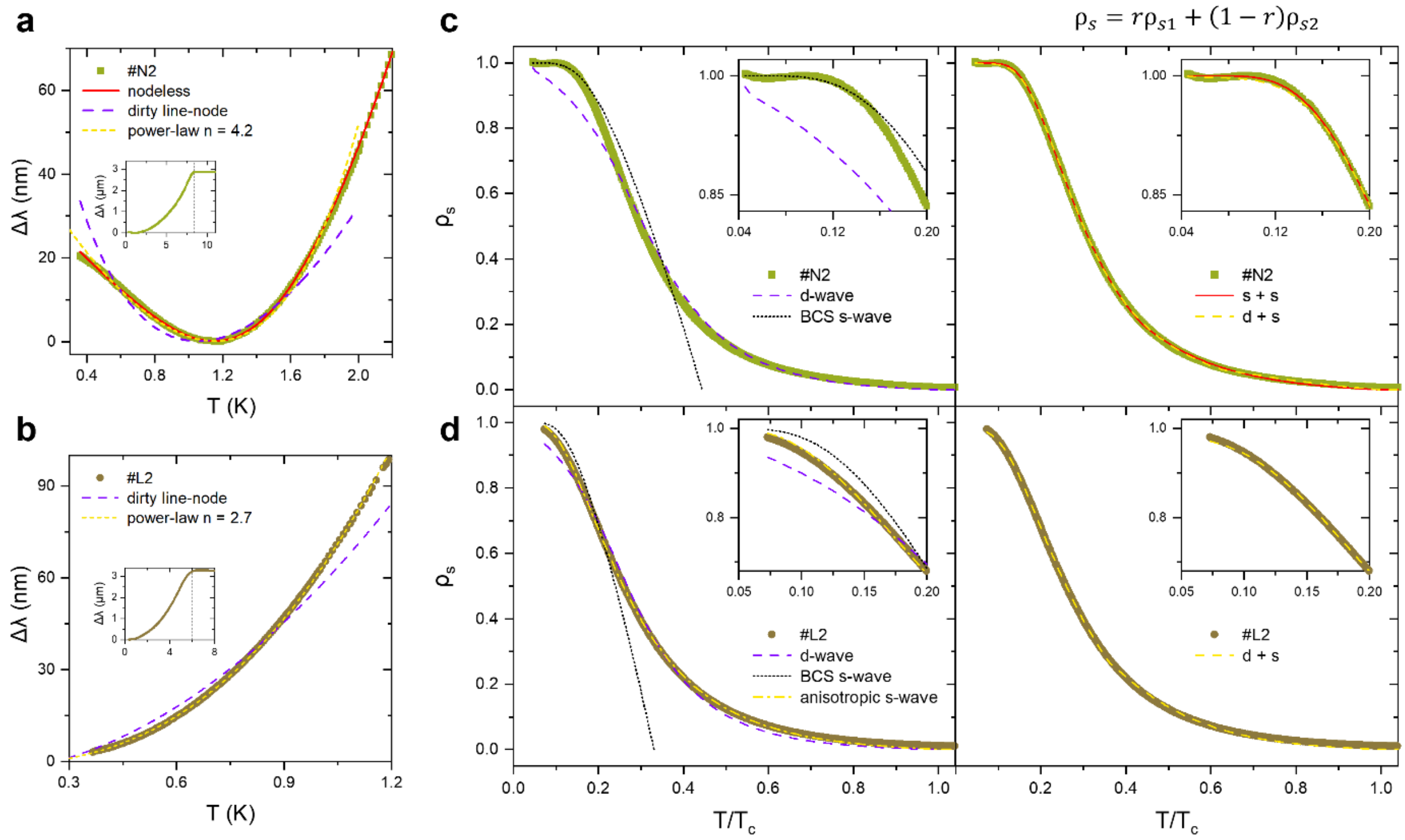

**Figure 4: London penetration depth and superfluid density for another set of Nd- (#N2) and La- (#L2) samples. (a-b)** Low-temperature fit of $\Delta\lambda$, insets show the full transition. The onset of transitions are $T_c = 8$ K for #N2 and $T_c = 6$ K for #L2. For #N2 (**a**), a similar low-temperature upturn is observed at below 1.2 K; such an upturn is absent in #L2 (**b**). $\Delta\lambda(T \ll T_c)$ is fitted to the nodeless exponential equation and power-law equation ($1 \leq n \leq 2$ for dirty line-node vs free $n$). Exponential equation traces #N2 data better; power-law equation with $n = 2.7$ best traces #L2 data. Best-fit parameters are listed in Table S1. **(c-d)** Global fit of $\rho_S$ for #N2 (**c**) and #L2 (**d**), where insets show the enlarged view at $T/T_c$ < 0.2. $\rho_S$ data is fitted to theoretical $\rho_S$ of single $d$-wave, BCS $s$-wave with variable $T_c$, anisotropic $s$-wave gap, and two gaps of different $T_c$. All best-fitted parameters are listed in Table S2.

**Superfluid density and multigap pairing**

Another crucial question on the pairing nature of the newfound infinite-layer nickelate superconductor family is whether they preserve multiband superconductivity like iron-pnictide[51] or a dominant single-band superconductivity like in cuprates[4]. While the low-temperature variation of the in-plane penetration depth $\Delta\lambda(T \ll T_c)$ and $\rho_S(T \ll T_c)$ discount a single $d_{x^2-y^2}$-wave pairing picture especially for the Nd-nickelate, to get a comprehensive picture of the pairing symmetry, we perform global fitting of the normalized superfluid density across the entire transition up to $T_c$. The zero-temperature in-plane London penetration depth $\lambda(0)$ of the Nd-sample is estimated using the Ginzburg-Landau equations based on the critical magnetic fields $H_{c1}$ and $H_{c2}$ (Figure S3) as described in the **Methods** section. We obtained a value of $\lambda(0) = 294 \pm 15$ nm and a coherence length $\xi(0) = 4.6 \pm 0.5$ nm. The calculated penetration depth is within the estimations from theoretical reports and a recent experimental study.[52,53] We use this value of $\lambda(0)$ to calculate the superfluid density $\rho_S$ for all samples. Regardless, we show in Figure 6 that the fitting results are qualitatively consistent over a range of $\lambda(0)$ values. For the Nd-based samples, the paramagnetic contribution is removed from the penetration depth $\lambda(T)$ data by dividing with $\sqrt{\mu}$ obtained from the low temperature fit, as shown in Figure S1.

Figure 3(c-d) and Figure 4(c-d) show the global fitting of the normalized superfluid density $\rho_S$ as a function of the normalized temperature $T/T_c$. The transition temperature $T_c$ is estimated as the temperature at which $\rho_S$ approximately reaches zero. The values of $T_c$ are 8.5K, 8K, 5.5K and 5K, respectively, for #N1, #N2, #L1 and #L2. We fit the superfluid density data to the theoretical $\rho_s$ (see **Methods** section) with a *d*-wave gap: $\Delta_d(\varphi, T) = \Delta_s(T)\cos(2\varphi)$ and a BCS weak-coupled $s$-wave gap: $\Delta_s(T) = \delta_{sc} k_B T_c \tanh\left\{\frac{\pi}{\delta_{sc}}\sqrt{a\left(\frac{\Delta C}{C}\right)\left(\frac{T_c}{T}-1\right)}\right\}$ with $T_c$ as the fitting parameter. The ratio of the gap magnitude at T = 0 K to $k_B T_c$, $\delta_{sc} = \frac{\Delta(0)}{k_B T_c}$, and the specific heat jump $\Delta C/C$ at $T_c$ are free parameters for $d$-wave fit and taken as $\delta_{sc} = 1.76, \Delta C/C = 1.43$ for BCS $s$-wave. The best-fit parameters are shown in Table S2. It is evident that the global fit of $\rho_s$ to a single $d$-wave gap fails for both Nd-nickelate and La-nickelate, especially at low temperatures. The BCS $s$-wave gap does not fit the data well either.

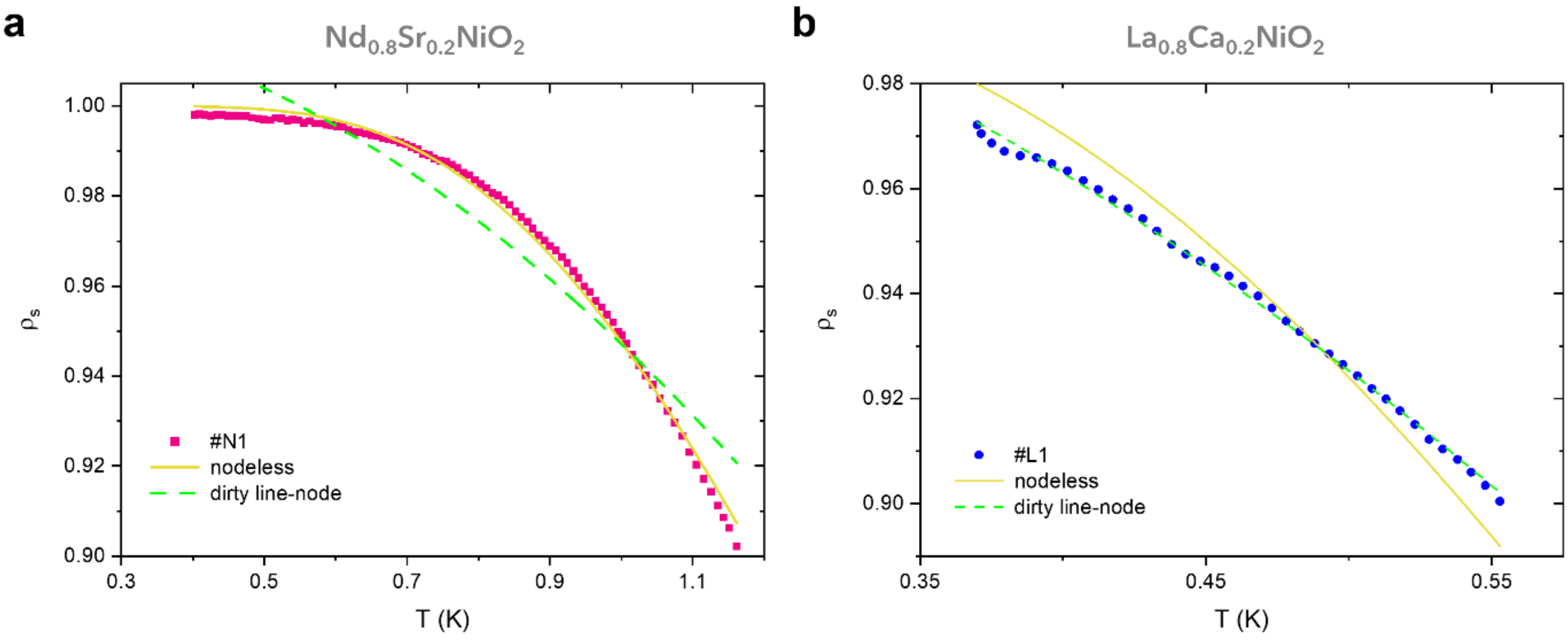

**Figure 5: Low-temperature fit of the superfluid density $\rho_S = [\lambda(0)/\lambda(T)]^2$ ($\propto$ phase stiffness $[1/\lambda(T)]^2$) at $\lambda(0) = 294$ nm.** Best fitted parameters are shown in Table 1.

While Nd-based samples (Figures 3c and 4c) show a flat region in $\rho_S$ at low temperatures, which agrees well with a nodeless gap, La-based samples do not show a flat region in $\rho_S$ down to the lowest temperature, which signifies an anisotropic nodal order parameter. Therefore, we explored the degree of anisotropy in the order parameter by fitting $\rho_S$ to an anisotropic $s$-wave of the form $\Delta_{anis} = \Delta_s(1 + \epsilon \cos 4\theta)$ where $0 \leq \epsilon \leq 1$ is a measure of the anisotropy; $\Delta_{anis}(\epsilon = 0)$ is the same as isotropic nodeless $s$-wave gap and $\Delta_{anis}(\epsilon = 1)$ is similar to a nodal $d$-wave gap. For the Nd-based samples, the best fit is obtained with zero anisotropy $\epsilon = 0$ with the same parameters as the isotropic $s$-wave (fitted parameters are given in Table S2). For the La-based sample, however, $\rho_s$ is fitted better with an anisotropic $s$-wave of nonzero $\epsilon = 0.37$ for #L1 and $\epsilon = 0.36$ for #L2. It is important to note that this anisotropic model is similar to the $(d + is)$-wave gap proposed in the theoretical calculations with $t - J - K$ model[35] since the fits of $\rho_s$ are sensitive to the magnitude of the superconducting gap magnitude only, but not its phase.

The obtained small values of gap magnitudes $\delta_{sc}$ and the long tail in $\rho_s$ data near $T_c$ suggest the possibility of a multigap scenario. Different multigap pairing scenarios can be used in the global fitting of the normalized superfluid density $\rho_S$. Since $d$-wave pairing is proposed in many theoretical calculations, we consider a two-gap model which includes both $d$-wave and $s$-wave gaps. For a two-gap model, individual superfluid densities add up as $\rho_s = r\rho_{s1} + (1 - r)\rho_{s2}$, where $r$ is a weight factor that determines the relative contribution from each gap[54]. We allow the transition temperature for the first gap ($T_{c1}$) to be a fitting parameter while $T_{c2}$ is a constant at the temperature at which $\rho_s$ approximately reaches zero as mentioned earlier. We have presented the most physical parameters found in each scenario in Table S2, and the fitted curves are shown in Figure 3(c-d) and Figure 4(c-d). In the case of the two-gap model, one of the gaps is expected to be larger than the weak-coupling limit $\delta_{sc}$ (1.76 for BCS $s$-wave and 2.14 for $d$-wave) and the other is expected to be smaller. For the Nd-based samples, $s + s$ pairing fits the data best for both #N1 and #N2. For example, #N1 has a larger $s$-wave gap $\delta_{sc} = 2.1 > 1.76$ at $T_{c1} = 4\text{K}$ with fraction $r = 0.53$ and another smaller $s$-wave gap $\delta_{sc} = 0.5$ at $T_{c2} = 8.5\text{K}$. For the La-based nickelates, we found $s + d$ pairing fits the $\rho_S$ data well with an $s$-wave gap $\delta_{sc} = 1.5 \pm 0.2$ at $T_{c1} = 1.8$ K of fraction $r = 0.46$ and a $d$-wave gap $\delta_{sc} = 2.2 \pm 0.5$ at $T_{c2} = 5.5$ K for sample #L1.

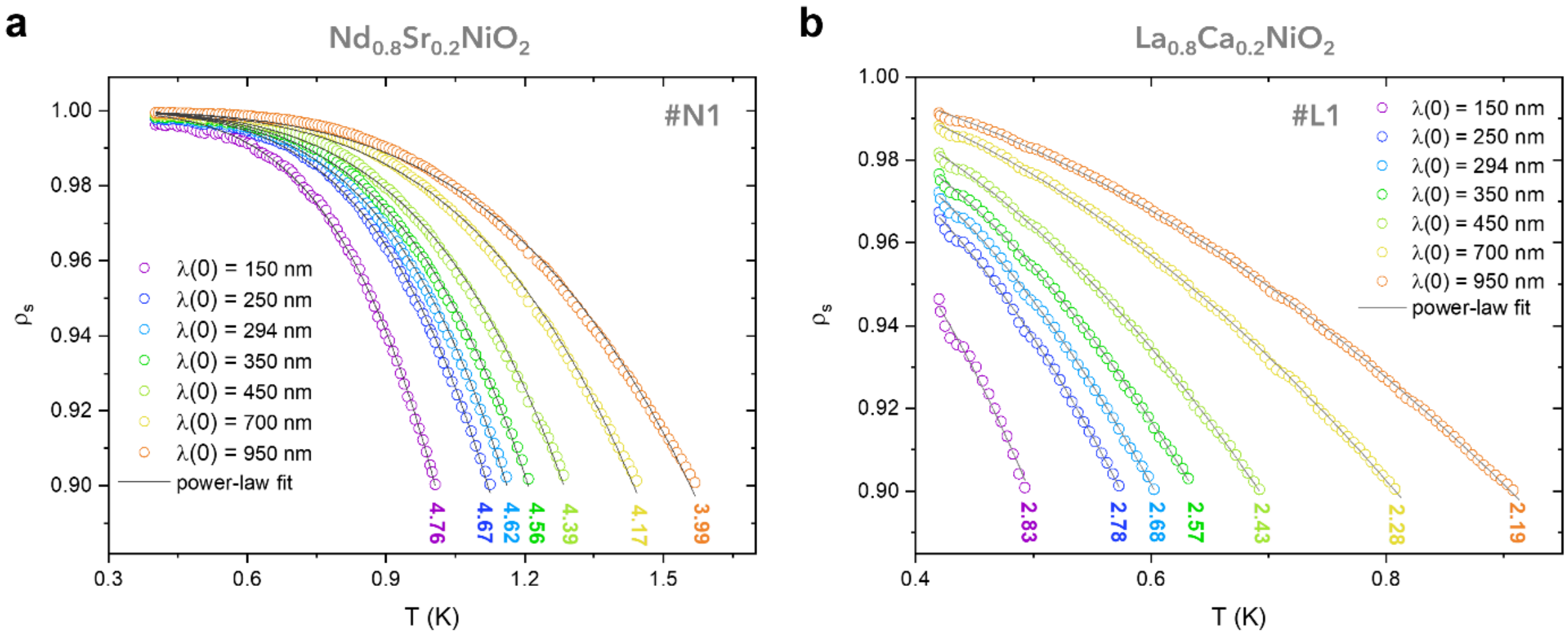

**Figure 6: Low-temperature fit of the superfluid density $\rho_S = [\lambda(0)/\lambda(T)]^2$ with a range of $\lambda(0)$ values.** The best-fitted parameter power $n$ from the power-law equation is annotated below the curve and shown in Table 2. Nd-nickelate shows a flat region in the low-temperature limit of phase stiffness and a larger deviation from nodal behaviour.

### Crossover from two-dimensional to three-dimensional superconducting state

In all the samples, the fitted specific heat jump $\Delta C/C$ is much smaller than the BCS weak-coupling value. This is a consequence of the long tail behaviour near $T_c$. The discussion in the supplementary information SI (D) has ruled out the favour of a wide transition as the cause. Such long-tail behaviour was seen previously in the quasi-1D superconductor $Tl_2Mo_6Se_6$ and explained as a dimensional crossover behaviour which arises from the coupling between the 1D superconducting chains[55]. With decreasing temperature, the superfluid density stays relatively flat after $T_c$ and rapidly rises at a lower crossover temperature after 3D phase coherence is established, which results in a 1D-to-3D dimensional crossover. In the superconducting infinite-layer nickelate of our present work, a similar crossover from two-dimensional to three-dimensional (2D-to-3D) superconducting states might be in play. The rare-earth $5d$ – Ni $3d$ orbital hybridization will form an interstitial-$s$ orbital which allows an extended $s$-wave gap to exist[35,56,57]. Below $T_c$, superconductivity first occurs in the two-dimensional $NiO_2$ plane, then at a lower temperature, 3D phase coherence between $NiO_2$ and rare earth spacer planes through interstitial-$s$ orbital is established, leading to a 2D-to-3D crossover. Such a dimensional crossover hypothesis is recently verified in an angular dependent upper critical field study.[45]

Our results have demonstrated that a single $d_{x^2-y^2}$-wave pairing cannot be the full picture in the infinite-layer nickelate superconductor family. While this experimental observation deviates from several early theoretical calculations[30–32], it is important to acknowledge that despite the isostructural and isoelectronic analogies[5,20], distinct electronic features were shown between nickelates and cuprates[14,16,37,56–59]. Recently, an upper critical field measurement revealed a large Pauli-limit violation in all directions in the La-nickelate,[42] which may be related to a possible stabilization of spin-triplet $p$-wave pairing on top of the $d$-wave pairing as predicted theoretically due to the influence of the Ni $d_{z^2}$ flat-band.[60] In addition, the isotropic nature[46] places Nd-nickelate to be more similar to the high-$T_c$ iron-based superconductor with nodeless multiband superconductivity[61]. In comparison, La-nickelate likely hosts an anisotropic nodal gap such as $(d+s)$- or $(d+p)$-wave in contrast to a nodeless gap, possibly $(d+is)$-wave, in the Nd-counterpart. On the other hand, the role of strain in the epitaxial thin-film form of nickelate superconductor further complicates the picture, as in-plane compressive stress is significantly larger in La- than Nd-nickelate thin-film. It is unable to be determined whether bulk nickelate[28] carries a distinct pairing due to the strain effects and a possibly different Ni $3d_{x^2-y^2}$ – O $2p$ orbitals hybridization. Furthermore, a crossover from two-dimensional to three-dimensional superconducting states can be interpreted from the observation of a long tail in the superfluid density near $T_c$ for both Nd- and La- infinite-layer nickelates, which likely reflects the roles of rare-earth spacer layer and epitaxial compressive stress. Our report presents a first investigation on the superconducting pairing symmetry of this family of superconductors, and should spur other experimental works and theoretical discussion.

## Methods

### Sample growth and preparation

The optimally doped ($x = 0.2$) perovskite nickelate $R_{1-x}A_xNiO_3$ ($\mathrm{R = La, Nd, A = Ca, Sr}$) thin film was grown on a $TiO_2$-terminated (001) $SrTiO_3$ (STO) substrate using a pulsed laser deposition (PLD) technique with a 248-nm KrF excimer laser. The deposition temperature for neodymium-based and lanthanide-based samples was set to 600°C and 576°C, respectively. The oxygen partial pressure $P_{O_2}$ was kept at 150 mTorr. The laser fluence on the target surface was 2.2 Jcm$^{-2}$ and 2.5 Jcm$^{-2}$ for the Nd-based nickelate and La-based nickelate, respectively. After deposition, the samples were annealed for 10 min at deposition temperature and 150 mTorr and then cooled down to room temperature at a rate of 8°C/min. Figure S4 shows the *in-situ* reflection high-energy electron diffraction pattern along the [100] crystallographic directions and *ex-situ* atomic force microscopy of the perovskite nickelate thin films after deposition.

For $CaH_2$ topotactic reduction, the sample was embedded with about 0.1 g of $CaH_2$ powder wrapped in aluminium foil and then placed into the PLD chamber. Using the PLD heater, the wrapped sample was heated to 340 – 360°C at a rate of 25 °C/min and kept for 60 – 80 minutes, and then cooled down to room temperature at a rate of 25 °C/min. No *in-situ* crystalline STO capping layer is introduced for all samples in this study. For the thickness of the infinite-layer thin film, Nd-based was consensual to be stabilized only up to ~10 nm, and hence we grew samples of 8 – 9 nm thick. La-based was reported to be stabilized at thickness ~17 nm,[37] and a larger coherent lattice thickness was calculated from the XRD Laue fringes. We grew 15 nm thick La-based films.

**Transport and magnetic measurement**

The wire connection for the electrical transport measurement was made by Al ultrasonic wire bonding. The transport measurements were performed using a Quantum Design Physical Property Measurement System. $T_{c,0}$ is defined as the temperature at which the resistivity reaches zero. The magnetic measurements were performed using a Quantum Design Superconducting Quantum Interference Device Magnetometer. Sample magnetic susceptibility was measured at $H \parallel c$ and the substrate diamagnetic background signal of ~$10^{-8}$ emu was subtracted before the calculation of volume susceptibility of nickelate thin film, as reported in Ref.[27]. Agreement in $T_c$ between resistivity, susceptibility and penetration depth measurement suggests a strong phase homogeneity across the entire film.

**Cross-sectional scanning transmission electron microscopy**

A focused ion beam (FIB) instrument (FEI Versa 3D for #N1, FEI FIB Scios for #L1), operated at 30 kV, was used to prepare cross-sectional lamellas of the nickelate thin films. Subsequently, cleaning at 2 kV was performed to remove any amorphous surface layer. The scanning transmission electron microscopy (STEM) characterization was conducted on two JEM-ARM200F (JEOL) microscopes, both operated at 200 kV, equipped with a cold field emission gun and a Cs-probe aberration corrector. High-angle annular dark-field (HAADF) images were acquired using inner and outer collection semi-angles of about 70 and 280 mrad, respectively, with a convergence semi-angle of about 30 mrad. A radial Wiener filter was used to reduce the background noise in the HAADF images. During HAADF imaging, the focus was tuned to the film-substrate interface. The electron microscopy investigations were conducted around three and seven months after sample synthesis for #N1 and #L1, respectively.

**London penetration depth measurement**

The change in London penetration depth, $\Delta\lambda(T)$ as a function of temperature, is measured using a homemade tunnel-diode-oscillator (TDO) based setup operating at 26 MHz with a resolution of around 0.01 Hz and low drift[43]. The rectangular cuboid sample, which is the infinite-layer nickelate thin film on STO substrate, is cut down to have a square basal plane of dimensions ~1 × 1 $mm^2$ and mounted on a sapphire rod using a thin layer of GE varnish. This is then mounted inside a Helium-3 cryostat and oriented inside the TDO coil such that $H_{ac} \parallel c$-axis to measure the change in the in-plane penetration depth $\Delta\lambda_{ab}$ given that $\Delta f \propto \Delta\lambda_{ab}$ (noted as $\Delta\lambda$ in the main text). The AC field inside the coil is less than 40 mOe which is smaller than the lower critical field of the sample. A bilayer mumetal jacket outside the cryostat shields external fields. The cryostat is able to cool the sample down to 0.35 K. A pre-calibrated Cernox 1030 temperature sensor from Lakeshore mounted at the bottom of the sapphire rod measures sample temperature. Heaters and sensors at various points allow for precise temperature control of all components. The sample temperature is varied while keeping the temperature of TDO components constant, and the change in frequency of the oscillator [$\Delta f_{ss}(T)$] (due to the change in susceptibility of the coil caused by the changing penetration depth of the sample inside it) is measured. The $\Delta f_{bg}(T)$ for the STO substrate is measured separately (together with other background signal) after dissolving the nickelate thin film by dipping in 1% HCl for 5 minutes. This background data [$\Delta f_{bg}(T)$] is then subtracted from the sample measurement to obtain an accurate frequency shift due to the thin film alone [$\Delta f(T) = \Delta f_{ss}(T) - \Delta f_{bg}(T)$]. The change in penetration depth of the sample is directly proportional to this change in frequency and is obtained as $\Delta\lambda(T) = G \cdot \Delta f(T)$, where $G$ is a calibration factor dependent on the sample and coil geometries [2]. The system is calibrated using 99.9995% pure Aluminium single crystal, which is a well-known non-local BCS superconductor. $G$ estimated using this technique might have an error of up to ~10% due to the irregularities in sample shapes. Regardless, $G$ is temperature independent and does not affect the discussion interpreted from the fitting in the penetration depth against exponential or power-law fitting models. Further details on the technique were reported previously[43,44].

The low-temperature in-plane London penetration depth $\Delta\lambda(T)$ was fitted to the (1) exponential BCS $s$-wave model $\Delta\lambda(T) \propto \left[\sqrt{\frac{\pi\Delta(0)}{2k_B T}} \exp\left(-\frac{\Delta(0)}{k_B T}\right)\right] + Y_0$, (2) clean line-node $\Delta\lambda(T) \propto T + Y_0$ and dirty line-node model $\Delta\lambda(T) \propto \frac{T^2}{T^*+T} + Y_0$,[50] or for both, $\Delta\lambda(T) \propto T^m + Y_0$ where $1 \leq m \leq 2$, and (3) power-law behavior $\Delta\lambda(T) \propto T^n + Y_0$ of arbitrary $n$. The free parameter $Y_0$ is introduced in the fitting equations to account for an unknown constant offset as the experimental $\Delta\lambda(T)$ is not measured down to 0 K (note: $\Delta\lambda(T) = \lambda(T) - \lambda(0)$). The normalized superfluid density ($\rho_s$) data is calculated from the change in penetration depth as,[54], $\rho_s(T) = \frac{\lambda^2(0)}{\lambda^2(T)} = \left(\frac{\Delta\lambda(T)}{\lambda(0)} + 1\right)^{-2}$ after subtracting the Curie-Weiss contribution from the measured $\Delta\lambda(T)$ as shown in Figure S1. The penetration depth at absolute zero, $\lambda(0)$ is calculated by solving equations,[54]

$$H_{c2}(0) = \sqrt{2}\kappa\, H_c(0)$$

$$H_{c1}(0) = \frac{H_c(0)}{\sqrt{2}\kappa}\, ln\, \kappa$$

$$H_{c1}(0) = \frac{\phi_0}{4\pi\, \lambda^2(0)}\, ln\, \kappa$$

$H_{c1}(0)$ and $H_{c2}(0)$ are the lower and upper critical magnetic fields, $\kappa = \lambda(0)/\xi(0)$ is the Ginzburg-Landau parameter, and $\phi_0 = 2.07 \times 10^{-7}$ Oe cm$^2$ is the magnetic flux quantum.

Using $H_{c1}(0) = 79$ Oe extrapolated from the magnetic measurements (Figure S3) and $H_{c2}(0) = 15.5$ T from resistivity measurements (Figure S3), we estimated $\lambda(0)$ to be $(294 \pm 15)$ nm and $\kappa = 64$. The value of in-plane coherence length $\xi(0) = 4.61$ nm obtained from these calculations agrees well with the previous report[46].

To interpret the superconducting gap profile, we performed a global fit of the calculated $\rho_s(T)$ data to the theoretically calculated superfluid densities using different gap models. We calculate the theoretical $\rho_s$ in the local limit for a two-dimensional Fermi surface with the expression,[54,62]

$$\rho_s = 1 + 2\int_0^{2\pi} \frac{d\varphi}{2\pi} \int_0^{\infty} \frac{\partial F}{\partial E} d\epsilon$$

where $\epsilon$ is the normal state quasi-particle energy, $E = \sqrt{\epsilon^2 + \Delta^2(T)}$ is the Bogulibov quasi-particle energy, and $F$ is the Fermi function given by $F = (exp(E/k_B T) + 1)^{-1}$. $\Delta(T)$ is the temperature-dependent superconducting gap. For an isotropic $s$-wave, we estimate the gap as[63]

$$\Delta_s(T) = \delta_{sc} k_B T_c \tanh\left\{\frac{\pi}{\delta_{sc}} \sqrt{a\left(\frac{\Delta C}{C}\right)\left(\frac{T_c}{T} - 1\right)}\right\}$$

where $\delta_{sc} = \Delta(0)/k_B T_c$ is the ratio of the magnitude of the gap at T = 0K to $k_B T_c$, $\Delta C/C \equiv \Delta C/\gamma T_c$ is the specific heat jump at $T_c$, and $a = 2/3$ is a constant. We use $\delta_{sc}$, $\Delta C/C$ and $T_c$ as fitting parameters. For a $d$-wave gap, this gap is modified with an angular dependence as $\Delta_d(\varphi, T) = \Delta_s(T)\cos(2\varphi)$. For two-gap models, individual superfluid densities add up as $\rho_s = r\rho_{s1} + (1 - r)\rho_{s2}$, where $r$ is a weight factor that decides the relative contribution from each gap. The low-temperature fit of the superfluid density $\rho_S = [\lambda(0)/\lambda(T)]^2$ ($\propto$ phase stiffness $[1/\lambda(T)]^2$) shown in Fig. 5 and Fig. 6 is performed with the following equations: (1)

exponential behaviour (nodeless): $\rho_s = 1 - \sqrt{\frac{\pi\Delta(0)}{2k_B T}} \exp\left(-\frac{\Delta(0)}{k_B T}\right)$, (2) dirty line-node: $\rho_s = \beta - \frac{\alpha T^2}{T+T^*}$, and (3) power-law $\rho_s = \beta - AT^n$.

## Authors contribution

A.A. conceived the project. A.A., E.E.M.C., L.E.C. and S.K.S. designed the experiments. L.E.C., S.W.Z., Z.T.Z. and X.M.D. synthesized the infinite-layer nickelate thin films. L.E.C. conducted the electrical and magnetic susceptibility measurements and analyzed the data. Z.Y.L., S.P. and Z.S.L. provided experimental support. Z.T.Z. measured the XRD data. S.K.S. performed the penetration depth measurement and analyzed the data with L.E.C, E.E.M.C. and A.A. The electron microscopy and spectroscopy investigations were analyzed by Z.Y.L. and conducted by P.N. (#N1) and T.H., J.D., P.A.v.A. (#L1). L.E.C., S.K.S., E.E.M.C. and A.A. wrote the manuscripts with input from all authors.

## Acknowledgement

We acknowledge useful discussion with Alexander A. Golubov, Harold Y. Hwang, Frank Lechermann, Karsten Held, Liang Si, Julia Mundy, Chunhui Rita Du, Swee Kuan Goh and King Yau Yip. This research is supported by the Ministry of Education, Singapore, under its MOE Tier 2 grant (Grant no. MOE-T2EP50121-0018). S.W.Z. and A.A. acknowledge earlier support by the Agency for Science, Technology, and Research (A*STAR) under its Advanced Manufacturing and Engineering (AME) Individual Research Grant (IRG) (A1983c0034). The authors would also like to acknowledge the Singapore Synchrotron Light Source (SSLS) for providing the facility necessary for conducting the research. The laboratory is a National Research Infrastructure under the National Research Foundation (NRF) Singapore. We acknowledge funding support from the European Union's Horizon 2020 research and innovation programme under grant agreement No 823717 - ESTEEM3.

## Data availability

The authors declare that the main data supporting the findings of this study are available within the article and its Supplementary Information files. Extra data are available from the corresponding author upon request.

**Table 1:** Best-fitted data at the low-temperature limit of the normalized superfluid density $\rho_s(T)$ ($\propto$ phase stiffness $[1/\lambda(T)]^2$) shown in **Fig. 5**.

| Sample | Nodeless exponential | Dirty line-node |
|---|---|---|
| $Nd_{0.8}Sr_{0.2}NiO_2$<br>#N1 | $\chi^2 = 3.2 \times 10^{-6}$<br>$\frac{\Delta(0)}{k_B} = (4.633 \pm 0.006)$ K | $\chi^2 = 5.6 \times 10^{-5}$<br>**Fail to fit**<br>$\alpha/T^* = 0.08$ K$^{-2}$ |
| $La_{0.8}Ca_{0.2}NiO_2$<br>#L1 | $\chi^2 = 3.4 \times 10^{-5}$<br>**Fail to fit**<br>$\frac{\Delta(0)}{k_B} = (2.108 \pm 0.008)$ K | $\chi^2 = 9.6 \times 10^{-7}$<br>$T^* = 1000$ K<br>$\alpha/T^* = 0.4$ K$^{-2}$ |

**Table 2:** Best-fitted data at the low-temperature limit of the normalized superfluid density $\rho_s(T)$ calculated using a range of $\lambda(0)$ values shown in **Fig. 6**.

| Sample | Power-law fit | | |
|---|---|---|---|
| | $\lambda(0)$ | $n$ | $\chi^2$ |
| $Nd_{0.8}Sr_{0.2}NiO_2$ #N1 | 150 nm | 4.76 | $1.2 \times 10^{-6}$ |
| | 250 nm | 4.67 | $6.8 \times 10^{-7}$ |
| | 294 nm | 4.62 | $6.9 \times 10^{-7}$ |
| | 350 nm | 4.56 | $7.0 \times 10^{-7}$ |
| | 450 nm | 4.39 | $1.1 \times 10^{-6}$ |
| | 700 nm | 4.17 | $1.3 \times 10^{-6}$ |
| | 950 nm | 3.99 | $1.6 \times 10^{-6}$ |
| $La_{0.8}Ca_{0.2}NiO_2$ #L1 | 150 nm | 2.83 | $2.4 \times 10^{-6}$ |
| | 250 nm | 2.78 | $5.6 \times 10^{-7}$ |
| | 294 nm | 2.68 | $3.9 \times 10^{-7}$ |
| | 350 nm | 2.57 | $3.0 \times 10^{-7}$ |
| | 450 nm | 2.43 | $2.3 \times 10^{-7}$ |
| | 700 nm | 2.28 | $1.9 \times 10^{-7}$ |
| | 950 nm | 2.19 | $1.7 \times 10^{-7}$ |